\documentclass[twocolumn, trackchanges]{aastex701}

\begin{document}

\title{DKIST Unveils Four-Lobed Stokes U Profiles in a Magnetically Complex Sunspot Penumbra}

\author[orcid=0000-0000-0000-0091]{Ryan J. Campbell}
\affiliation{Astrophysics Research Centre, Queen's University Belfast, Belfast, Northern Ireland, BT7 1NN, United Kingdom}
\email[show]{ryan.campbell@qub.ac.uk}
 
\author[orcid=0000-0001-5459-2628]{Sarah A. Jaeggli}
\affiliation{National Solar Observatory,
22 Ohia Ku Street, Pukalani, HI 96768, United States of America}
\email{sjaeggli@nso.edu}

\begin{abstract}

We present DKIST spectropolarimetric observations of a sunspot penumbra revealing spatially coherent four-lobed Stokes $U$ profiles that persist after polarimetric crosstalk correction. The observed Stokes vectors are well reproduced by stratified inversions containing gradients in velocity and magnetic parameters, including line-of-sight variation of the magnetic azimuth. Forward synthesis experiments in which the magnetic azimuth is artificially held constant with optical depth substantially modify the Stokes $U$ morphology while leaving Stokes $I$, $Q$, and $V$ comparatively unchanged, indicating that azimuthal variation is an important contributor to the observed linear polarisation signatures within the tested one-component atmospheres. We explored degeneracies using both two-component atmospheres and one-component constant-azimuth models: the former generally converged toward pathological solutions, while the latter provided a plausible degenerate explanation for much of the observed four-lobed Stokes $U$ morphology. A supervised machine-learning classifier applied to Stokes $V$ profiles shows that the associated magnetic complexity is spatially confined to the penumbra–light-bridge boundary where the four-lobed Stokes $U$ profiles and reverse-polarity fields are observed. These results identify four-lobed Stokes $U$ profiles as a potential spectropolarimetric signature of complex line-of-sight magnetic structuring in sunspot penumbrae, providing new observational constraints on fine-scale penumbral magnetic topology.

\end{abstract}

\section{Introduction} 
The structure of the penumbral magnetic field and its relation to plasma flow models has been the subject of intense debate in solar physics. Morphologically, the penumbral magnetic field can be understood as consisting of spines, with relatively strong and vertical magnetic fields, and intraspines, with weaker and more horizontal fields. Total pressure balance implies that intraspines must be elevated with respect to spines due to their reduced magnetic pressure \citep[see review by][]{borrero2011}. High–spatial-resolution vector magnetometry has long indicated that the penumbral magnetic field exhibits substantial fine-scale variation \citep{Lites1993}. Earlier interpretations of anomalous sunspot polarisation signals invoked gradients in velocity and magnetic field orientation along the line of sight, including even postulating differential azimuthal twists, to explain integrated circular polarisation signatures \citep{Makita1986}. Modern radiative magnetohydrodynamic simulations have since reached a level of realism such that the observed penumbral fine structure appears to be reproduced, predicting strong vertical gradients in magnetic field orientation associated with magnetised flows and returning flux \citep{rempel2012}.

Despite these advances, a central limitation in testing penumbral models has been the difficulty of isolating unambiguous spectropolarimetric diagnostics of vertical magnetic structuring. Studies based on Hinode and GREGOR observations have demonstrated that key signatures of penumbral magnetism, such as multi-lobed Stokes $V$ profiles associated with reverse-polarity magnetic fields (RPMFs), are strongly suppressed or misclassified when spatial smearing is present, even when spectral resolution and polarimetric sensitivity are high \citep{franz2013}. Image deconvolution analyses have further suggested that unresolved structure can mask the true prevalence of complex magnetic topology in the penumbra \citep{ruizcobo2013}. The highest-resolution pre-DKIST observations revealed radially aligned lanes of reversed polarity embedded within the penumbra \citep{scharmer2013}.

The Daniel K. Inouye Solar Telescope (DKIST), with its 4-m aperture, enables diffraction-limited spectropolarimetric observations in the visible at spatial scales of order $0\farcs1$, combined with high spectral resolution and polarimetric sensitivity \citep{campbell2023, rast2021, rimmele2020}. In this study, we present a close analysis of a sunspot penumbra observed by DKIST with the Visible Spectropolarimeter \citep[ViSP;][]{ViSP}. 

\begin{figure*}
    \centering
    \includegraphics[width=\linewidth]{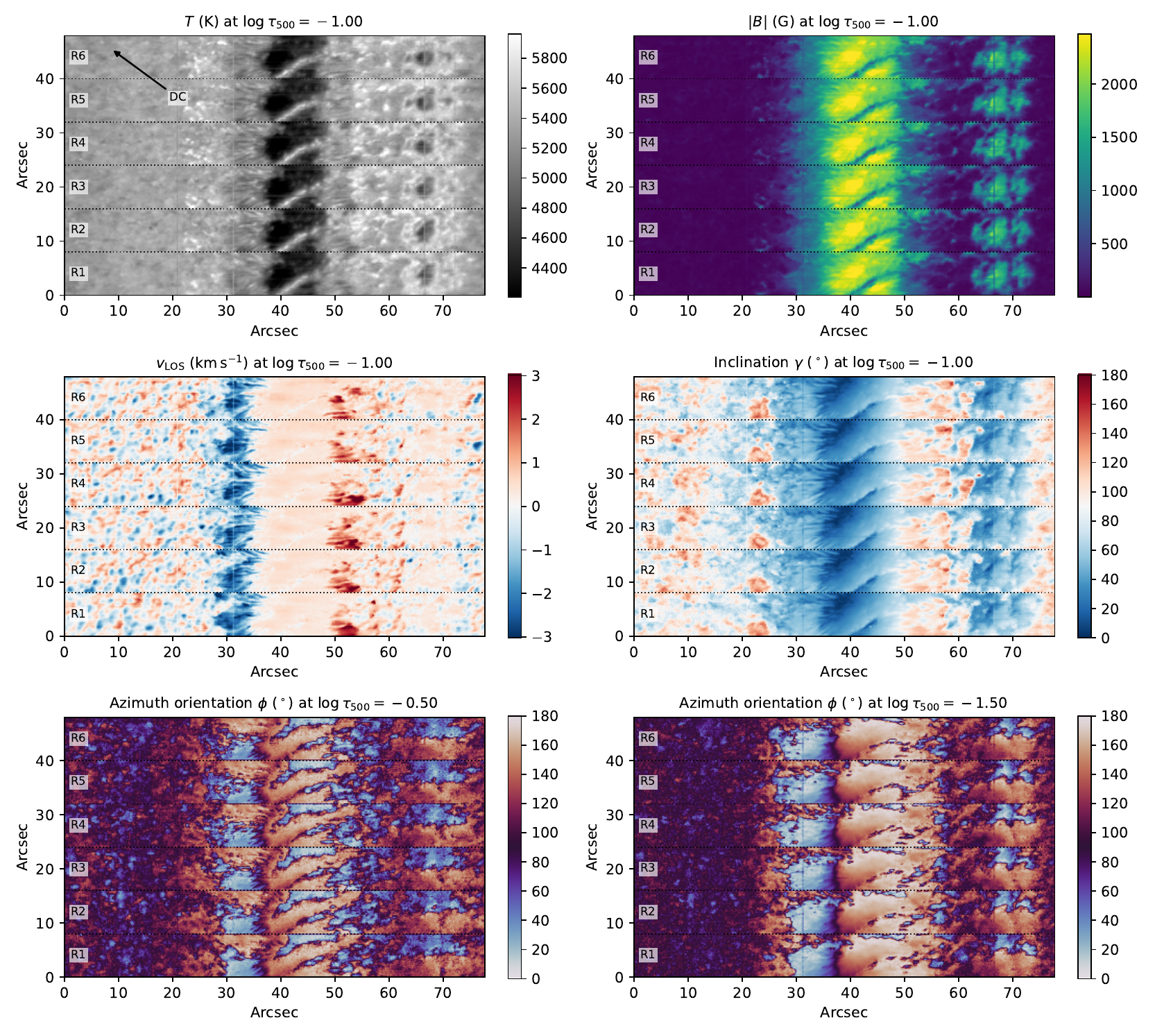}
    \caption{Full-field inversion maps for the DKIST/ViSP sunspot dataset. LTE inversion results for temperature, magnetic field strength, line-of-sight velocity, and magnetic field inclination are shown at $\log\tau = -1.0$. The modulo of the azimuth is shown at two depths ($\log\tau = -0.5$ and $\log\tau = -1.5$). All six raster maps (labelled R1-R6) are stacked along the scan direction to form the composite maps shown here. Consequently, the vertical axis represents the cumulative spatial coordinate of the sequence rather than the coordinate within a single raster frame. The black arrow indicates the approximate location of disk centre (DC).}
    \label{fig:maps}
\end{figure*}

\begin{figure*}
    \centering
    \includegraphics[width=\linewidth]{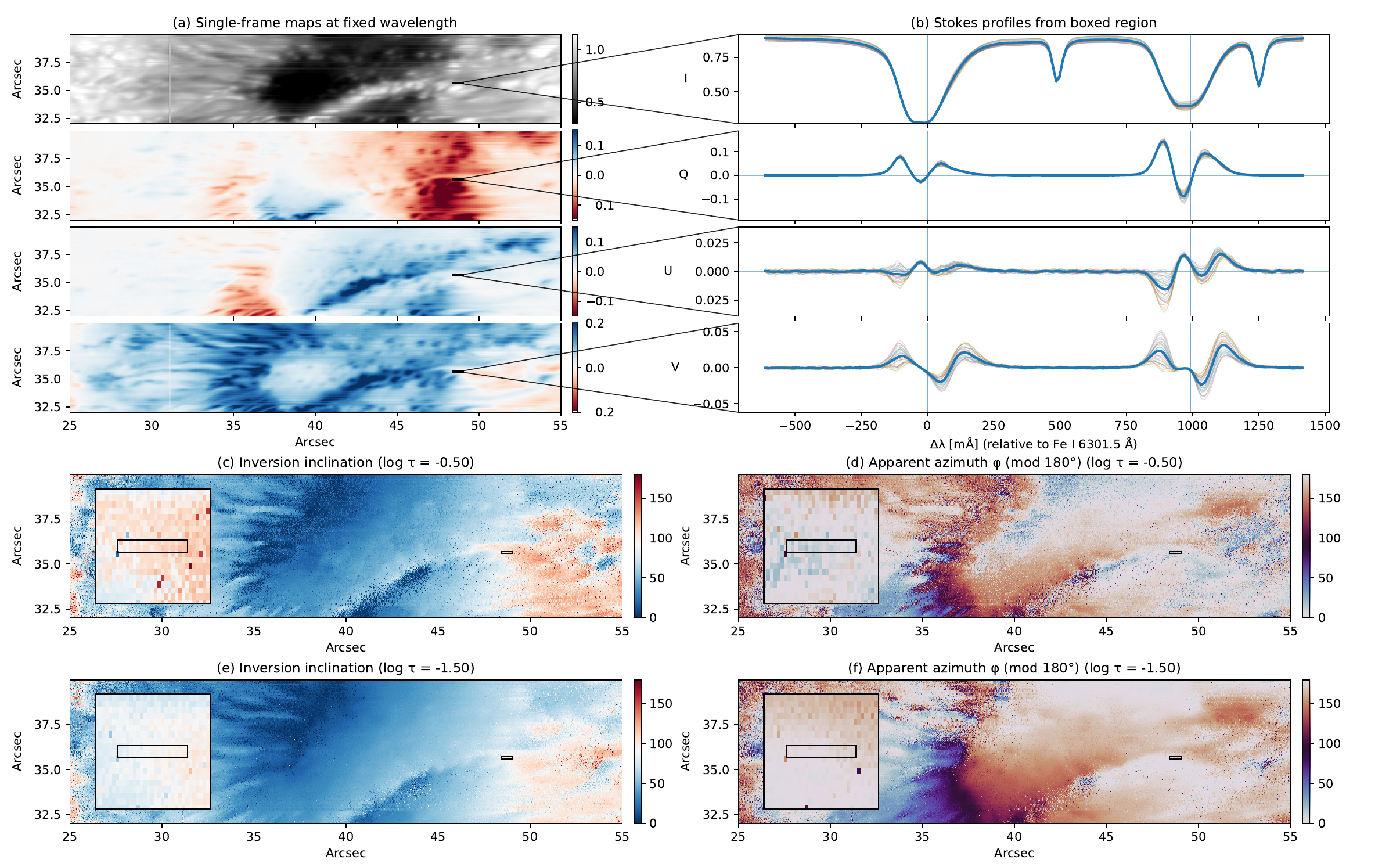}
    \caption{
DKIST/ViSP observations reveal spatially coherent, multi-lobed Stokes $U$ profiles in a sunspot penumbra, while the majority of linear polarisation profiles remain morphologically normal.
(a) Single-wavelength maps of Stokes $I$, $Q$, $U$, and $V$ from a single ViSP raster scan across the sunspot. Stokes $I$ is shown at a continuum wavelength, Stokes $Q$ and $U$ are shown at the line core, and Stokes $V$ is shown in the blue wing of Fe~I~6302.5~\AA. The boxed region highlights a small penumbral segment used for the profile analysis in panel (b).
(b) Stokes profiles extracted from the small boxed region: thin lines show individual pixel profiles, while the thick solid line denotes the mean profile over the region. The profiles are plotted as a function of wavelength offset relative to Fe~I~6301.5~\AA. The multi-lobed structure observed in Stokes $U$ is spatially coherent across neighbouring pixels. Panels (c)--(f) show maps of inclination and azimuth angles at two representative optical depths ($\log\tau = -0.50$ and $-1.50$.) The location of the four-lobed Stokes U signals is highlighted by the small box in each panel, as in the upper panels. The left-most insets show a zoomed-in $1\arcsec$ by $1\arcsec$ region centered on the small boxed region.}
    \label{fig:fig1}
\end{figure*}

\begin{figure*}
    \centering
    \includegraphics[width=\linewidth]{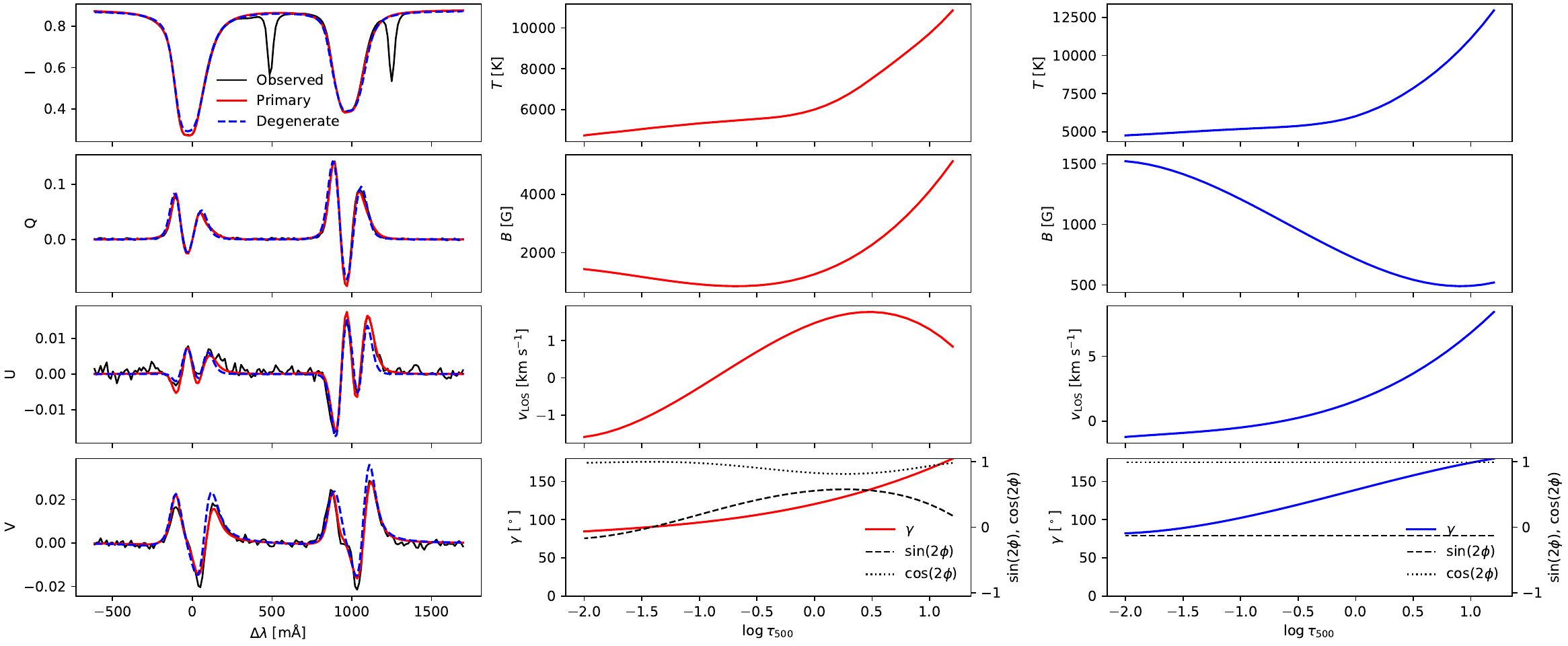}
    \caption{
Comparison between depth-dependent and constant-azimuth inversion solutions for a representative penumbral pixel exhibiting four-lobed Stokes $U$ profiles. \textit{Left column}: observed Stokes $I$, $Q$, $U$, and $V$ profiles (black) together with the best-fit depth-dependent azimuth solution (red) and a plausible one-component constant-azimuth degenerate solution (blue dashed). \textit{Middle column}: inferred atmospheric stratifications for the Primary solution (i.e. with depth-dependent azimuth), including temperature, magnetic field strength, line-of-sight (LOS) velocity, magnetic inclination, and azimuth (represented by $\sin(2\phi)$ and $\cos(2\phi)$). \textit{Right column}: corresponding atmospheric stratifications for the constant-azimuth degenerate solution. }
    \label{fig:StokesU_with3lobedV}
\end{figure*}

\section{Observations}
On 07 August 2025 between 19:33 and 20:45 UTC observations of an active region (NOAA 14169) were acquired with the ViSP at the DKIST during its third cycle of observations. The observing sequence was designed to obtain narrow, repeated scans of a sunspot. The selected slit width was $0{\,}.{\!\!}{\arcsec}0536$ with a slit step of $0{\,}.{\!\!}{\arcsec}054$ and $150$ slit positions per raster map, giving a map of $8{\,}.{\!\!}{\arcsec}1\ \times 76{\,}.{\!\!}{\arcsec}14$. The cadence between the commencement of a given raster map to the start of the next was $14.4$ minutes, and $6$ raster maps were recorded in total. One arm of the ViSP captured a spectral region which includes the magnetically sensitive photospheric Fe I line pair at $6301.5$~$\mathrm{\AA}$ and $6302.5$~$\mathrm{\AA}$ (with effective Land\'e g-factors of $1.67$ and $2.5$, respectively). In this study we focus our analysis on the Fe I doublet. The spatial sampling along the slit is $0{\,}.{\!\!}{\arcsec}0304$/pixel for the arm which recorded the Fe I $6300$~$\mathrm{\AA}$ spectral region.  The linear dispersion in this spectral region is $12.76$~$\mathrm{m\AA}$/pixel. The observations were acquired away from disk centre corresponding to a viewing angle of $\mu=0.8$.

\section{Methodology}

\subsection{Polarimetric crosstalk removal}\label{sect:crosstalk}
Initially, we applied the method of \citet{crosstalk1992} to remove instrumental cross-talk from Stokes $I$ into Stokes $Q$, $U$, and $V$. This correction removes continuum polarisation offsets: after its application, the median continuum signal in Stokes $Q$, $U$, and $V$ is consistent with zero, and no residual polarisation is detected at the wavelengths of the $\mathrm{O}_2$ telluric lines. We then applied the ensemble method described by \cite{crosstalk}, which models residual polarimetric mixing as a single field-of-view–averaged Mueller transformation. Residual cross-talk between linear and circular polarisation states was estimated using strongly polarised pixels by fitting an elliptical retardance Mueller matrix describing rotations in Stokes space that minimise correlations between Stokes $V$ and $Q,U$. This correction accounts for residual $Q/U \leftrightarrow V$ mixing introduced by imperfect retardance calibration. This method has also been employed in recent DKIST analyses of plage regions \citep{joao2023}. To avoid biasing the solution with complex penumbral profiles, the ensemble used to estimate this rotation was restricted to umbral pixels using an intensity threshold. The multi-lobed Stokes $U$ profiles that are the focus of this analysis are already present in the pre-corrected data and remain after the umbra-constrained ensemble correction. This result is stable for reasonable variations of the umbral intensity threshold.

\subsection{Inversions}\label{sect:maps}

In order to infer the atmospheric stratifications that reproduce the observed Stokes profiles, we perform inversions using the DeSIRe code \citep{desire}. Maps of the parameters are shown in Fig.~\ref{fig:maps}. Each pixel was inverted ten times using randomised initial values for magnetic field strength, $B$, line-of-sight (LOS) velocity, $v_{\mathrm{LOS}}$, magnetic inclination, $\gamma$, and magnetic azimuth, $\phi$, and the solution exhibiting the minimum $\chi^2$ was retained. This approach reduces sensitivity to initial conditions and mitigates convergence to local minima. Elemental abundances were adopted from \citet{asplund}.
In DeSIRe, the one-dimensional models are perturbed at node-points at fixed equidistant atmospheric depths. Four inversion cycles were employed. In the first cycle, each parameter was perturbed with only one node, but in subsequent cycles increasing depth-dependent perturbation was introduced culminating in a maximum of six nodes in $B$, $\gamma$, and $v_{\mathrm{LOS}}$ and nine nodes in $T$ in the final cycle to allow for gradients required to reproduce asymmetric and multi-lobed Stokes profiles.

The wavelength scale was calibrated by comparison with a Fourier Transform Spectrometer (FTS) solar atlas. The inferred LOS velocities were post-processed by enforcing a zero mean velocity in magnetically quiet regions of the scan outside the sunspot.

The magnetic field inclination and azimuth returned by the inversions are defined in the observer’s LOS reference frame, as returned directly by the inversion code. No transformation into the local solar reference frame was applied in this work. The magnetic inclination $\gamma$ is defined relative to the observer’s LOS direction, while the magnetic azimuth $\phi$ describes the orientation of the transverse magnetic field component in the plane perpendicular to the LOS.

\subsection{Supervised classification}\label{sect:MLP}
To assess the spatial distribution of Stokes $V$ morphologies across the penumbra, we performed a supervised classification of Stokes $V$ profiles using a Multi-Layer Perceptron (MLP) neural network, following the general philosophy of the profile-based analysis of \cite{campbell2025}. Profiles were extracted from a wavelength window encompassing the Fe~I~6302.5~\AA\ line only and normalised by their maximum absolute amplitude to isolate profile shape from absolute polarisation strength. The classifier was trained exclusively on Stokes~$V$ profiles, as its primary purpose is to assess whether regions exhibiting complex linear polarisation signatures are systematically associated with three-lobed circular polarisation morphologies. The trained model was then applied to the full dataset to generate spatial maps of Stokes~$V$ profile classes, enabling a direct comparison with the locations of four-lobed Stokes~$U$ profiles.

A subset of profiles was manually labelled into a small number of morphology-based classes using an interactive graphical interface. These labels were used to train the MLP in a supervised manner according to a $70-15-15\%$ training-validation-test split. In particular, Stokes $V$ profiles were assigned to one of four classes: (i) positive-polarity, two-lobed profiles consistent with simple Zeeman splitting; (ii) two-lobed, positive-polarity profiles exhibiting pronounced magneto-optical effects; (iii) negative-polarity, two-lobed profiles (with or without magneto-optical signatures); and (iv) multi-lobed profiles displaying more complex morphology not obviously attributable to simple magneto-optical distortions. Stokes $V$ profiles belonging to class (i) and (ii) are very similar, except that in strong and inclined fields magneto-optical effects can produce additional lobes. \cite{campbell2025} referred to such profiles as ``double'' profiles because they clearly have the same dominant polarity as the sunspot umbra, but have two additional smaller lobes near the line center. Negative-polarity profiles exhibiting analogous morphologies were not separated into distinct subclasses because such profiles occurred only rarely within the dataset, preventing robust supervised training of an additional dedicated class.

The MLP classifier follows the methodology introduced in \citet{campbell2025}, and consists of two fully connected hidden layers with nonlinear Swish activation functions and dropout regularisation. The MLP was optimised using a weighted cross-entropy loss function to account for class imbalance. No spatial information was provided to the classifier during training, indicating that any observed spatial coherence arises from intrinsic structure in the spectropolarimetric signals rather than from imposed smoothing or contextual constraints. A hyperparameter sweep over network architecture and training parameters (including hidden-layer size, dropout fraction, learning rate, and batch size) was performed, and the best-performing model was selected according to macro-averaged F1 scores across all classes. Per-class F1 scores were also inspected to ensure a balanced performance across all classes. Once trained, the classifier was applied to the full dataset, assigning each pixel to the class with the highest posterior probability. The best model had a generalisation accuracy of $96\%$ and an F1-score of $94\%$ on the held-out test set.

\section{Results}

Figure~\ref{fig:fig1} presents single-wavelength maps of Stokes $I$, $Q$, $U$, and $V$ for one raster map, revealing a conventional sunspot morphology albeit with a light-bridge. However, within a confined penumbral segment highlighted in Fig.~\ref{fig:fig1}a, the Stokes $U$ signal exhibits an unusual morphology. As shown in Fig.~\ref{fig:fig1}b, Stokes $U$ profiles in this region display a spatially coherent, four-lobed morphology that persists across neighbouring pixels and is unlikely attributable to noise or unresolved lateral mixing within the spatial resolution element. These profiles are present in the pre-corrected data and persist after application of the ensemble crosstalk correction, indicating that they are not introduced by polarimetric mixing.

The corresponding Stokes $V$ profile is three-lobed. The profiles shown have already undergone polarimetric cross-talk correction, the signals are spatially coherent along penumbral lanes, and they appear in the same spatial locations between two raster maps. Most importantly, they are only visible on one side of the sunspot, whereas if they were the result of significant and systematic crosstalk we could expect to see similar profiles at the same slit-positions on both opposing penumbral regions. 

The lower panels of Fig.~\ref{fig:fig1} also show the inferred magnetic field inclination and apparent azimuth at two representative optical depth layers. In the highlighted penumbral region, the azimuth displays a change between $\log\tau = -0.50$ and $-1.50$. The median absolute azimuth difference within the boxed region is $11^\circ$ (computed using the shortest ambiguity-consistent separation), indicating a coherent rotation of the transverse field within the line-formation region.

The question naturally arises whether the observed Stokes profiles--most notably the spatially coherent, multi-lobed Stokes $U$ signals--are actually reproduced within a standard local thermodynamic equilibrium (LTE) inversion framework. Figure~\ref{fig:StokesU_with3lobedV} demonstrates that this is indeed the case. For a representative penumbral pixel drawn from the region highlighted in Fig.~\ref{fig:fig1}, an LTE inversion yields an excellent fit to the full Stokes vector, including the complex Stokes $U$ morphology. The inferred atmospheric stratifications exhibit physically plausible gradients in temperature, magnetic field strength, LOS velocity, and field orientation, without requiring discontinuities or extreme parameter values. The inversion retrieves depth-dependent structure in all atmospheric parameters. The three-lobed Stokes $V$ profile can be interpreted as an indication of significant inclination gradients that cross the $90^\circ$ polarity line at an optical depth to which the line is highly sensitive \citep{campbell2023}, which is indeed the case in the inversion solution. This confirms that the observed linear polarisation signatures are compatible with realistic one-dimensional atmospheric models.

\begin{figure*}
    \includegraphics[width=\textwidth]{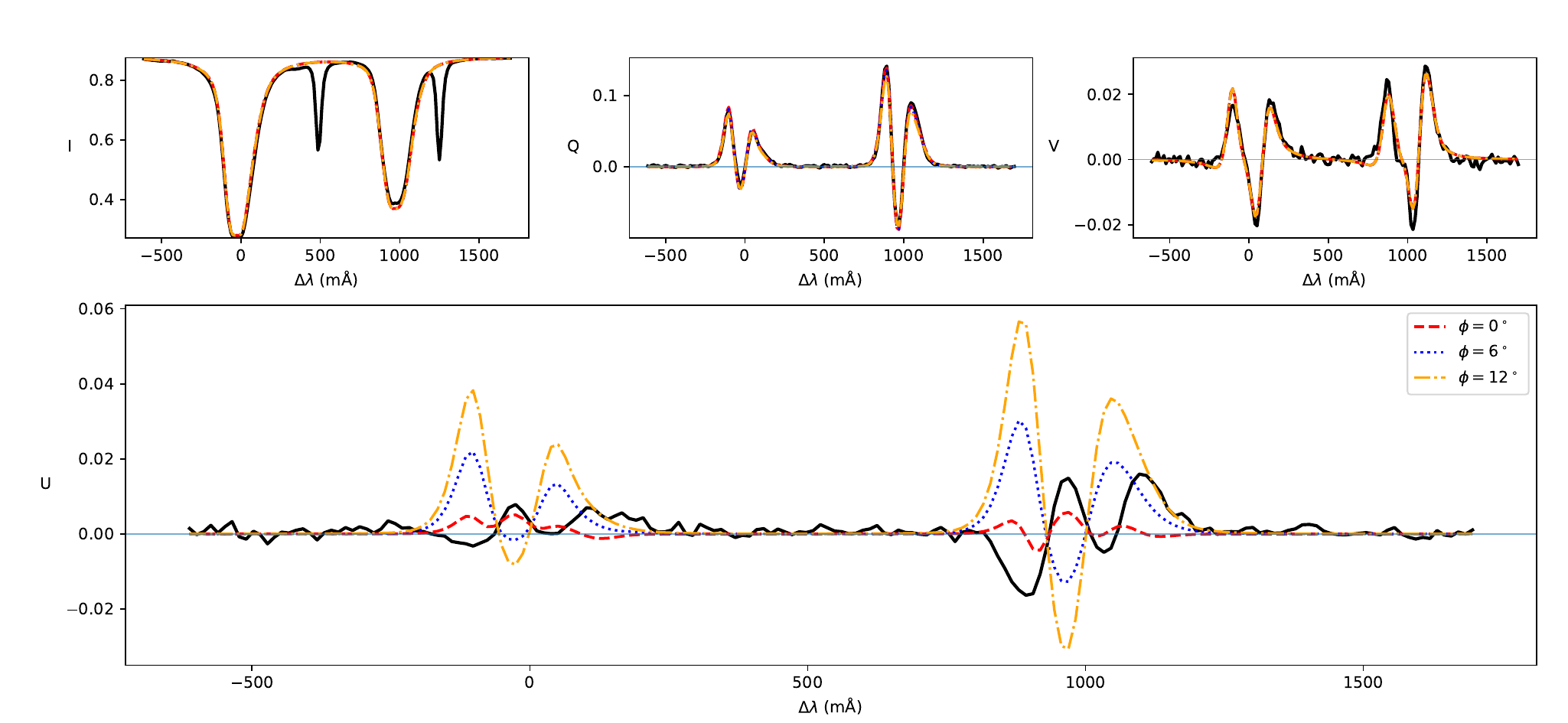}
    \caption{Forward synthesis tests isolating the role of magnetic azimuthal stratification. The black lines show the observed Stokes profiles for the same representative penumbral pixel exhibiting multi-lobed Stokes $U$ shown in Fig.~\ref{fig:StokesU_with3lobedV}. Coloured lines show forward syntheses from modified atmospheres in which the magnetic azimuth is forced to be constant with optical depth (values indicated in the legend), while all other atmospheric parameters are kept fixed at their inversion-retrieved stratifications. Stokes I, Q, and V remain unchanged under this modification, whereas the four-lobed Stokes $U$ morphology is not reproduced without a gradient in $\phi$.}
    \label{fig:forward_synthesis}
\end{figure*}

In an attempt to isolate the role of the magnetic azimuthal shear, we first performed forward synthesis tests in which the azimuth was artificially forced to be constant with optical depth, while all other atmospheric parameters were kept fixed at their inversion-retrieved stratifications. The adopted constant values correspond to the inversion-inferred azimuth at $\log\tau=-0.50$ ($\approx12^\circ$), at $\log\tau=-1.50$ ($\approx0^\circ$), and their mean ($6^\circ$), spanning the vertical azimuthal range recovered in the best-fit model.\footnote{The magnetic azimuth $\phi$ is defined modulo $180^\circ$ due to the intrinsic ambiguity of the transverse Zeeman effect. In the inversion, the azimuth is encoded via $\sin(2\phi)$ and $\cos(2\phi)$.} The resulting Stokes profiles are shown in Fig.~\ref{fig:forward_synthesis}. Under these modifications, Stokes $I$, $Q$, and $V$ remain largely unchanged, indicating that the temperature structure, magnetic field strength gradients, and LOS velocity gradients alone are sufficient to reproduce those components. However, the multi-lobed morphology of Stokes $U$ is not recovered in any of the constant-azimuth cases. This demonstrates that the vertical variation of the magnetic azimuth within the line-formation region was a required ingredient in this solution to produce the observed linear polarisation structure.

We additionally explored two-component inversions in which the magnetic azimuths of the two atmospheric components were allowed to differ while remaining constant with optical depth within each component. In the tested cases, the inversions either converged toward negligible secondary filling factors ($\lesssim 10^{-2}$) or required unrealistically large magnetic field strengths ($\gtrsim 10^{5}$~G) and LOS velocities ($\gtrsim 10^{2}$~$\mathrm{km\,s^{-1}}$) in the secondary component, without substantially improving the quality of the Stokes profile fits relative to the single-component depth-dependent azimuth solution. We therefore find no compelling evidence that unresolved multi-component structure alone can account for the observed multi-lobed Stokes $U$ profiles. Although the additional component could formally reproduce aspects of the four-lobed Stokes $U$ morphology, it did so primarily by acting as a highly shifted, pathological correction term with vanishing filling factor, rather than as a physically meaningful secondary magnetic atmosphere.

We additionally explored whether the observed four-lobed Stokes $U$ profiles could be reproduced within a one-component atmosphere while enforcing a magnetic azimuth that remains constant with optical depth. Figure~\ref{fig:StokesU_with3lobedV} shows the results of this test. The resulting inversion provides a plausible degenerate solution that reproduces much of the observed Stokes morphology without requiring LOS azimuthal variation. However, the degenerate solution achieves this through substantially different atmospheric stratifications, including a different magnetic field strength gradient and LOS velocity structure. Conversely, the inferred magnetic inclination stratifications remain broadly similar. The existence of this solution demonstrates that the interpretation of the four-lobed Stokes $U$ profiles is not unique within a one-component inversion framework.

\begin{figure*}
    \centering
    \includegraphics[width=\linewidth]{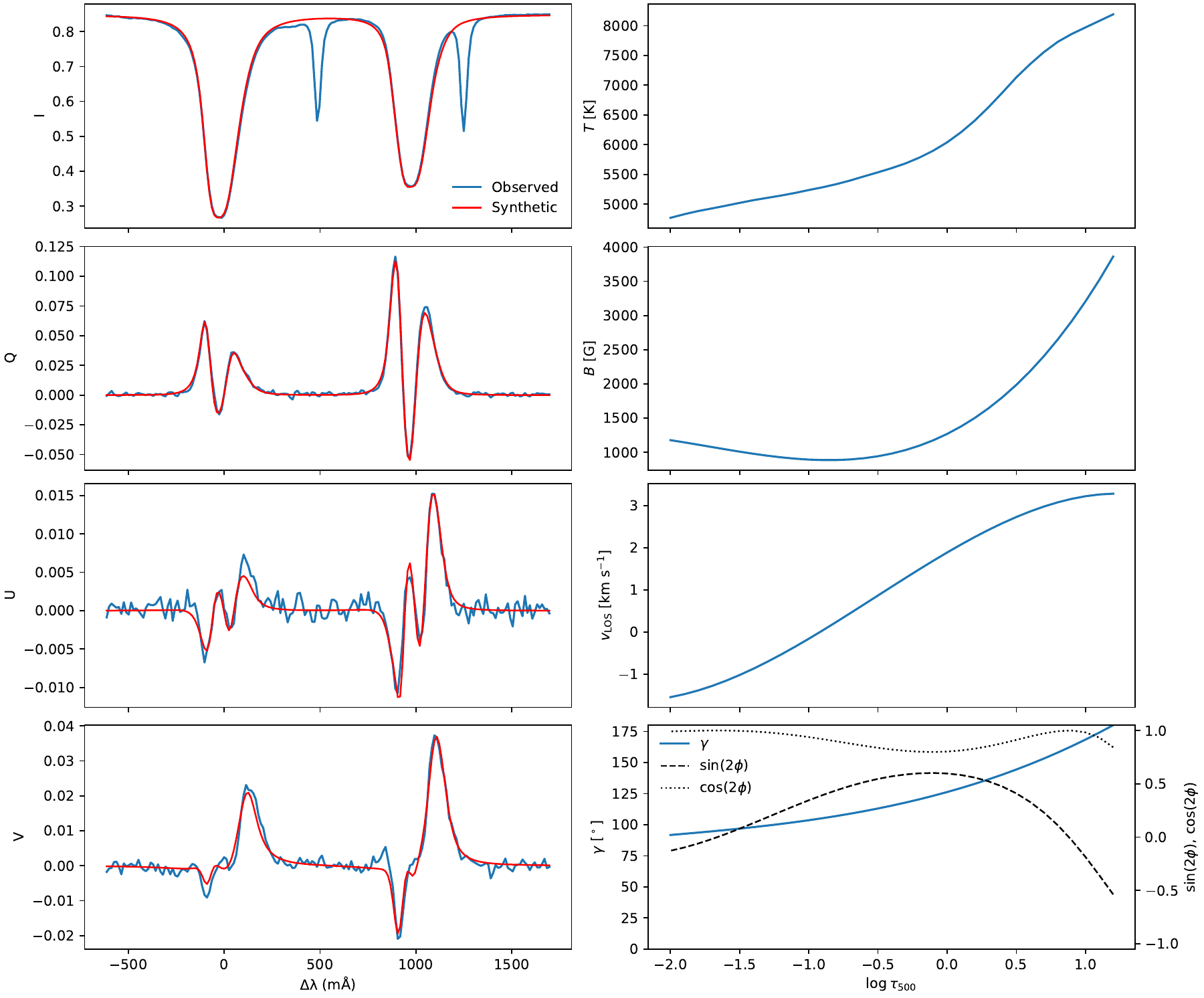}
\caption{
As in Fig.~\ref{fig:StokesU_with3lobedV}, but for a penumbral pixel exhibiting a four-lobed Stokes $U$ profile together with a comparatively simple two-lobed Stokes $V$ morphology. This example demonstrates that complex Stokes $U$ signatures are not restricted to regions displaying three-lobed Stokes $V$ profiles. The inversion shown here allowed the magnetic azimuth to vary with optical depth.}
    \label{fig:StokesU_withoutV}
\end{figure*}

\begin{figure*}
    \centering
    \includegraphics[width=\linewidth]{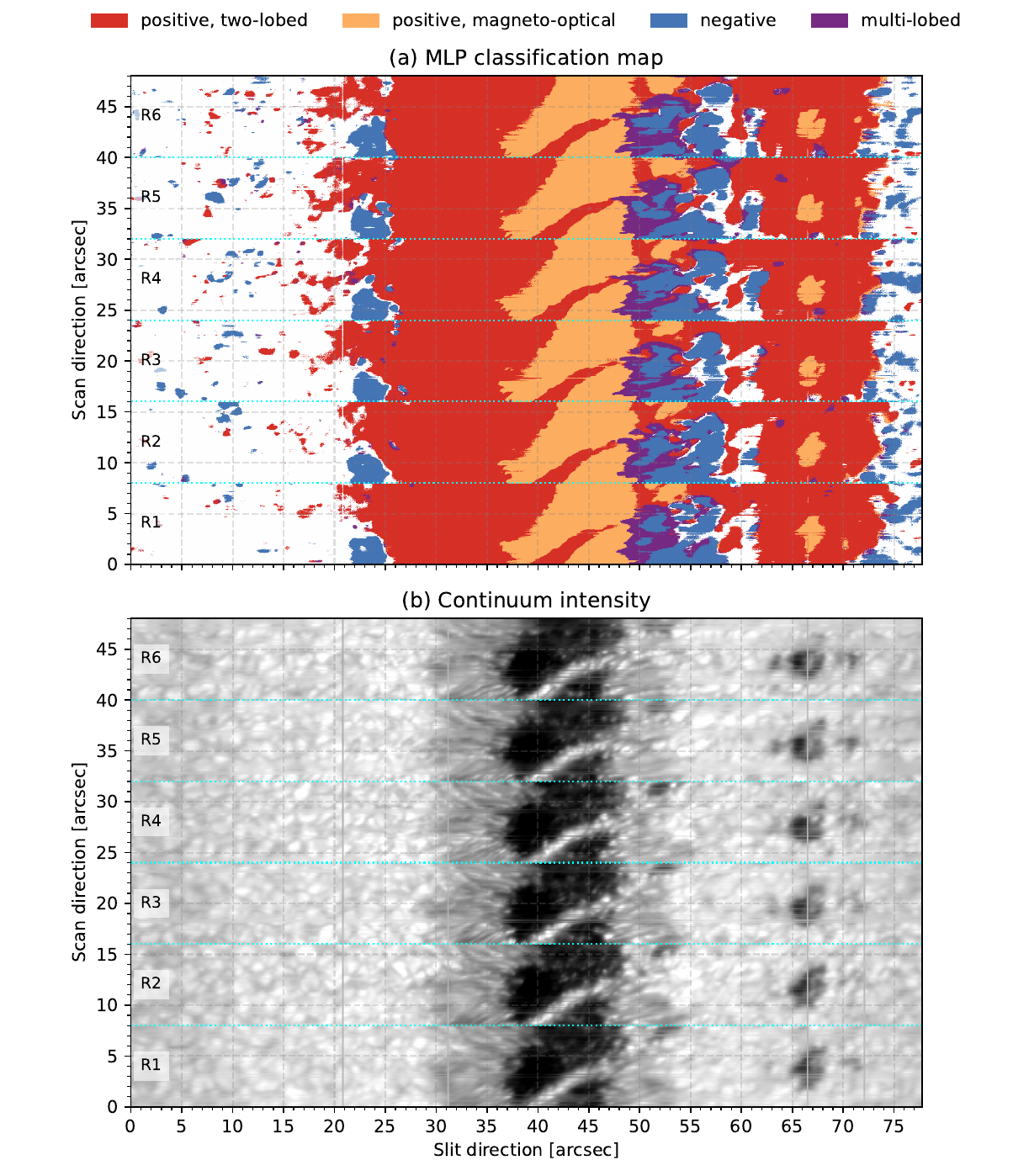}
\caption{Spatial distribution of Stokes $V$ profile morphologies and the corresponding continuum intensity across the full DKIST raster sequence. Panel (a) shows the supervised MLP classification map, where each pixel is classified solely from the Stokes $V$ spectral shape within the Fe~I~630.25~nm line and assigned to one of four morphology classes representing dominant-polarity two-lobed (\textit{red}), dominant-polarity magneto-optical (\textit{yellow}), reverse-polarity/minority-polarity (\textit{blue}), and multi-lobed (\textit{purple}) Stokes $V$ profiles. Panel (b) shows the corresponding continuum intensity map for spatial reference. The six raster scans, labelled R1 to R6, are stacked along the scan direction and separated by cyan dotted lines. The vertical and horizontal dashed lines are gridlines to aid visual inspection.}
    \label{fig:MLP}
\end{figure*}

Figure~\ref{fig:StokesU_withoutV} shows an additional example of a Stokes $U$ profile, located in an adjacent region of the penumbra as the example shown in Fig.~\ref{fig:fig1} and Fig.~\ref{fig:StokesU_with3lobedV}. In this example, the Stokes $V$ profile is two-lobed. The occurrence of four-lobed Stokes~$U$ profiles is spatially confined to the penumbra, where they coexist with a spatially coherent mixture of reverse-polarity two-lobed and three-lobed Stokes~$V$ profiles. Crucially, however, there is no obvious one-to-one correspondence between the presence of complex Stokes~$V$ morphologies and the occurrence of four-lobed Stokes~$U$ profiles at the pixel level. There are cases where pixels exhibiting simple, two-lobed Stokes~$V$ profiles nevertheless show complex Stokes~$U$ structure, while conversely not all pixels with multi-lobed Stokes~$V$ display anomalous Stokes~$U$ profiles. This lack of morphological and spatial correspondence strongly disfavours a direct $V \rightarrow U$ crosstalk explanation, as linear leakage would be expected to preserve the basic lobe structure of the parent profile. Moreover, the confinement of the effect to the penumbra--rather than its ubiquitous appearance across regions of strong circular polarisation--argues against a purely instrumental origin.

The classification map of Stokes $V$ profile morphologies is shown in Fig.~\ref{fig:MLP}. Complex Stokes $V$ morphologies, including multi-lobed and reverse-polarity profiles, are not uniformly distributed but occur in spatially coherent patches within the penumbra. In contrast, the umbra and light-bridge regions are dominated by simpler two-lobed profiles, which show magneto-optical effects near the line core when the magnetic field is more inclined with respect to the LOS. At $\mu=0.8$, the magneto-optical effects are more prominent in the umbra than the light-bridge as the vertical umbral field is more inclined with respect to the LOS. Importantly, the regions exhibiting complex Stokes $V$ morphologies coincide with the penumbral sector where four-lobed Stokes $U$ profiles are observed. This spatial confinement argues against a global instrumental origin and is consistent with the presence of localised magnetic complexity associated with penumbral return flux.

In Fig.~\ref{fig:MLP}, in the \textit{left-most} penumbral regions and the light-bridge, the Stokes $V$ profiles are predominantly two-lobed. There is a significant reverse polarity patch outside the penumbra, which is also visible in the inclination maps provided in Fig.~\ref{fig:fig1}. On the \textit{right-most} penumbra, all four classes of Stokes $V$ profiles can be found. Here, there is a significant amount of RPMFs, evidenced both by the detection of negative polarity profiles and multi-lobed profiles. The existence of the multi-lobed profiles at the interface of the two polarities is physically consistent with a transition between the dominant polarity of the umbra and the RPMFs found in the penumbra. Finally, the classifications between raster maps are spatially coherent, similar but different according to solar evolution.

\section{Discussion}
When the magnetic azimuth varies along the line of sight, different atmospheric layers contribute linear polarisation signals with different orientations. We report the detection of spatially coherent four-lobed linear polarisation signatures for the first time to our knowledge. These profiles persist after careful, model-based polarimetric crosstalk correction, are spatially coherent along penumbral lanes, appear in the same spatial locations in more than one sequential raster, and can be reproduced within an LTE inversion framework. 
Forward-synthesis experiments in which the magnetic azimuth was artificially flattened while holding all other atmospheric parameters fixed demonstrated that the Stokes $U$ morphology is particularly sensitive to the azimuthal stratification within the tested atmospheres. However, additional one-component inversion experiments also revealed plausible constant-azimuth degenerate solutions capable of reproducing much of the observed Stokes morphology through substantially different atmospheric stratifications. The inferred magnetic inclination stratifications remained broadly similar between the solutions despite differences in magnetic-field-strength and velocity gradients, suggesting that some aspects of the magnetic geometry are more robustly constrained than others. Although similarly complex linear polarisation profiles may exist in observations from earlier facilities such as Hinode/SP, they have not been explicitly reported. This study may motivate a targeted re-examination of such datasets.

Because the observations were acquired away from disk centre ($\mu = 0.8$), the line of sight samples an inclined path through the penumbral atmosphere. Within the one-dimensional inversion framework, this LOS variation corresponds to a rotation of the transverse magnetic field along the sampled atmospheric column. Curved magnetic fields contribute to the Lorentz force through the magnetic-tension term,
\begin{equation}
\mathbf{F}_L =
\underbrace{-\nabla \left(\frac{B^2}{2\mu_0}\right)}_{\text{magnetic pressure}}
+
\underbrace{\frac{1}{\mu_0}(\mathbf{B}\cdot\nabla)\mathbf{B}}_{\text{magnetic tension}} .
\end{equation}
However, the present observations do not permit a direct quantitative evaluation of the full force balance in the penumbra. If the depth-dependent azimuth solution is adopted, this would suggest spatial variation in the magnetic-field direction that would contribute to magnetic tension in the Lorentz force, while remaining consistent with magnetohydrostatic equilibrium if these stresses are balanced by gas-pressure gradients.

The supervised classification of Stokes $V$ morphologies provides additional context. The MLP analysis reveals that the penumbral sector where the four-lobed Stokes $U$ profiles are found is also where circular polarisation profiles are most complex, containing both three-lobed and reverse-polarity Stokes $V$ signals. This spatial correspondence indicates that the four-lobed Stokes $U$ profiles arise within the most magnetically structured part of the penumbra, consistent with expectations for regions containing strong gradients or returning flux. The confinement of the effect to a specific penumbral sector further argues against a purely instrumental origin, since residual polarimetric mixing would be expected to correlate primarily with signal amplitude rather than with local magnetic morphology.

\section{Conclusions}
We report the detection of spatially coherent four-lobed Stokes $U$ profiles, to our knowledge for the first time, in a DKIST observation of a magnetically complex sunspot penumbra. The observations demonstrate that four-lobed Stokes $U$ profiles provide a new observational signature of complex line-of-sight magnetic structuring in sunspot penumbrae. Stratified inversions with depth-dependent azimuth reproduce these profiles naturally, but one-component constant-azimuth inversions can reproduce much of the morphology through different atmospheric stratifications, demonstrating that the interpretation is not unique within the tested inversion framework. Future multi-line observations and inversions may help further constrain the physical origin of these profiles and determine the extent to which they diagnose magnetic azimuthal shear, unresolved structure, or more general atmospheric complexity.

\begin{acknowledgments}
We thank the anonymous reviewer for comments that improved the manuscript. R.J.C would like to thank Maxim Kramar, Tetsu Anan, and David Kuridze for their help and expertise with experiment generation and execution via service mode operation of DKIST. R.J.C thanks Mihalis Mathioudakis for insightful discussions. 
This is UKRI-supported research under the Science and Technology Facilities Council (STFC) grant ST/X000923/1. The research reported herein is based in part on data collected with the Daniel K. Inouye Solar Telescope (DKIST), a facility of the National Solar Observatory (NSO). NSO is managed by the Association of Universities for Research in Astronomy, Inc., and is funded by the National Science Foundation. Any opinions, findings and conclusions or recommendations expressed in this publication are those of the author(s) and do not necessarily reflect the views of the National Science Foundation or the Association of Universities for Research in Astronomy, Inc.

\end{acknowledgments}

\begin{contribution}
R.J.C. conceived the project, performed the analysis, interpreted the results, and wrote the manuscript. S.J. provided expert guidance on cross-talk removal and feedback on the manuscript and physical interpretations.
\end{contribution}

\facilities{DKIST/ViSP \citep[][]{rimmele2020, ViSP}}

\software{DeSIRe \citep[][]{desire}, StokesClassifier \citep[][]{campbell2025}}

\bibliography{sample701}{}
\bibliographystyle{aasjournalv7}

\end{document}